\documentclass[11pt]{amsart}

\usepackage{amsfonts}
\usepackage{amsmath,amssymb,amsthm,enumerate,epsfig,graphicx}
\usepackage{epstopdf}
\usepackage{ifthen,latexsym,syntonly}
\usepackage{natbib}
\usepackage{rotating}
\usepackage{lscape}
\usepackage{subfigure}
\usepackage{float}
\usepackage{caption}
\usepackage{comment}
\usepackage{supertabular}
\usepackage{hyperref}
\usepackage{color}
\usepackage{multirow}
\usepackage{booktabs}
\usepackage{threeparttable}

\usepackage{color}
\usepackage{colortbl}

\theoremstyle{definition}

\allowdisplaybreaks

\begin{document}	

\title[Detecting sites with biased readings of $PM_{2.5}$ by HDGCM]{Bias detection of $PM_{2.5}$ monitor readings using hidden dynamic geostatistical calibration model}

\author[Yaqiong Wang]{Yaqiong Wang}
\address{Peking University,
	Peking,
	China.}
\author{Minya Xu}
\address{Peking University,
	Peking,
	China.}
\email{minyaxu@gsm.pku.edu.cn}
\author{Hui Huang}
\address{Sun Yat-sen University,
	Guangzhou,
	China.}
\author{Songxi Chen}
\address{Peking University,
	Peking,
	China.}

\begin{abstract}
	Accurate and reliable data stream plays an important role in air quality assessment. Air pollution data collected from monitoring networks, however, could be biased due to instrumental error or other interventions, which covers up the real pollution status and misleads policy making. In this study, motivated by the needs for objective bias detection, we propose a hidden dynamic geostatistical calibration (HDGC) model to automatically identify monitoring stations with constantly biased readings. The HDGC model is a two-level hierarchical model whose parameters are estimated through an efficient Expectation-Maximization algorithm. Effectiveness of the proposed model is demonstrated by a simple numerical study. Our method is applied to hourly $PM_{2.5}$ data from 36 stations in Hebei province, China, over the period from March 2014 to February 2017. Significantly abnormal readings are detected from stations in two cities.
\end{abstract}

\maketitle
\keywords{Hidden dynamic geostatistical calibration model; Bias detection; Expectation-Maximization algorithm; $PM_{2.5}$ data}
	
\section{Introduction}

In recent years, China has experienced severe regional air pollutions \citep{hu2010air}, because of rapid industrialization and alarming increase in energy consumption. Particulate matters with diameter smaller than 2.5 microns ($PM_{2.5}$) is one of the major pollutants and receives a lot of public attention. Lots of studies have shown that $PM_{2.5}$ is very harmful to human health through penetrating deeply into lung tissues and entering the bloodstream. For example, long-term exposure to $PM_{2.5}$ may cause lung cancer \citep{pope2002lung} or cardiorespiratory failure \citep{hoek2013long}, and regional $PM_{2.5}$ pollution may increase mortality rate (e.g. \cite{schwartz1996is}, \cite{shang2013systematic} and \cite{lelieveld2015contribution}). High concentrations of $PM_{2.5}$ also influence people's daily life such as outdoor activities and routine business (e.g. \cite{pope1995review} and \cite{chapko1976air}). 

The Chinese government identifies the needs for air quality assessment and emission control, and has built a large monitoring network all over the country since 2013. Now there are over 1,500 national stations in over 300 cities. Hourly readings of air pollutants are regularly recorded and directly transferred to China National Environmental Monitoring Center (CNEMC). To guarantee appropriate air pollution management, accurate and reliable data stream is pressingly required, otherwise serious health damages and economic implications would follow. Data quality of monitor readings, however, may vary across regions and time. It can be affected by instrumental failures, improper maintenance or certain interventions (e.g., using sprinklers and fog guns around monitoring stations). So far, the CNEMC simply picks up outliers by visual checking, which requires a lot of manpower and may not identify abnormal data promptly. Therefore, an efficient and objective way to detect inaccurate readings is urgently needed.

Methods to evaluate the reliability of air pollution data are limited in literature. In \cite{Fass2007Air}, a spatio-temporal calibration model for $PM_{10}$ data in Italy was developed to manage a heterogeneous air pollution monitoring network caused by instrumental biases, where some instruments were known to underestimate the ``true'' $PM_{10}$ level. In their method, dimension reduction was addressed by deterministic basis function from decomposition of the spatial covariance. However, such a technique failed to provide a sufficient description of a spatio-temporal process evolution \citep{huang2007model}.
Another work from \cite{ghanem2014effortless} claimed that the evidence of data manipulation could be found in the discontinuities of the probability density functions of air pollution index (API, usually determined by $PM_{10}$) around the threshold value of 100 for different Chinese cities ($API<100$ indicating ``blue-sky days''; otherwise, indicating ``non-blue-sky days"). This choice of threshold was too arbitrary, though informative in some cases, can not be directly applied to general pollutants and more complex pollution scenarios.

Due to the complex nature of environmental data, especially their underlying spatiotemporal structure, it is better to fully use the information from the nearby monitoring stations and time points to calibrate the readings of target site. Spatiotemporal statistical model, which considers both the spatial covariance and temporal dependence, is thus a good and natural choice \citep{cressie2015statistics}. \cite{huang1996spatio} introduced a widely used spatiotemporal model in environmental field, which included a dynamic random field with a separable spatiotemporal covariance structures. By referring the model as the hidden dynamic geostatistical (HDG) model, \cite{Calculli2015Maximum} generalized the HDG model by adding the effects from covariates. Although the generalized HDG model performed well in modelling air pollutant dynamics, it cannot be directly applied to calibration problems.

In our study, based on the generalized HDG model, we develop a hidden dynamic geostatistical calibration (HDGC) model to identify biased readings from monitoring stations. Particularly, in the HDGC model, we add a calibration component for each station to account for the station-specific effect. 
If certain calibration component is substantially lower than neighboring stations, it will send an alarm that those stations have abnormal readings. In addition to incorporating the calibration component, the HDGC model can also take covariates into consideration, and reveal spatial correlation structure and temporal dynamic mechanism in data. Therefore, we explore the association between $PM_{2.5}$ and other pollutants, meteorological variables, as well as geographical factors. To some degree, it helps provide insight of the formation mechanism of $PM_{2.5}$ pollution. In the study, we apply the modified HDGC model to $PM_{2.5}$ readings of 36 stations in Hebei, China, a province often suffering from severe air pollution where many heavy industry factories are deployed. 

The paper is organized as follows. In Section 2, we describe the data of the study region. In Section 3, we establish the HDGC model with calibration components, and introduce Kalman filter and expectation maximization (EM) algorithm for parameter estimation, following by variance covariance matrix for model parameters. Then we conduct a simulation study to validate the performance of our model in Section 4. We give a comprehensive interpretation of the results in Sections 5 and provide conclusions in Section 6. 

\section{Data Description}

In this study, we specially target Hebei province, as it is one of the most heavy air polluted areas in China. In the list of top ten polluted cities in China, more than half belong to Hebei province. This province is thus the main target of complaints regarding air pollution. Hebei province locates in north China with most of its central and southern parts lie within the North China Plain. The western part of Hebei rises to the Taihang Mountain, while the eastern part is bordered by Bohai Gulf. In the province, there locates many heavy industry enterprises, especially iron and steel manufactures. For years, the coal consumption in Hebei has been the highest among all provinces in China. 

\begin{figure}
	\centering
	\makebox{\includegraphics[width=8cm]{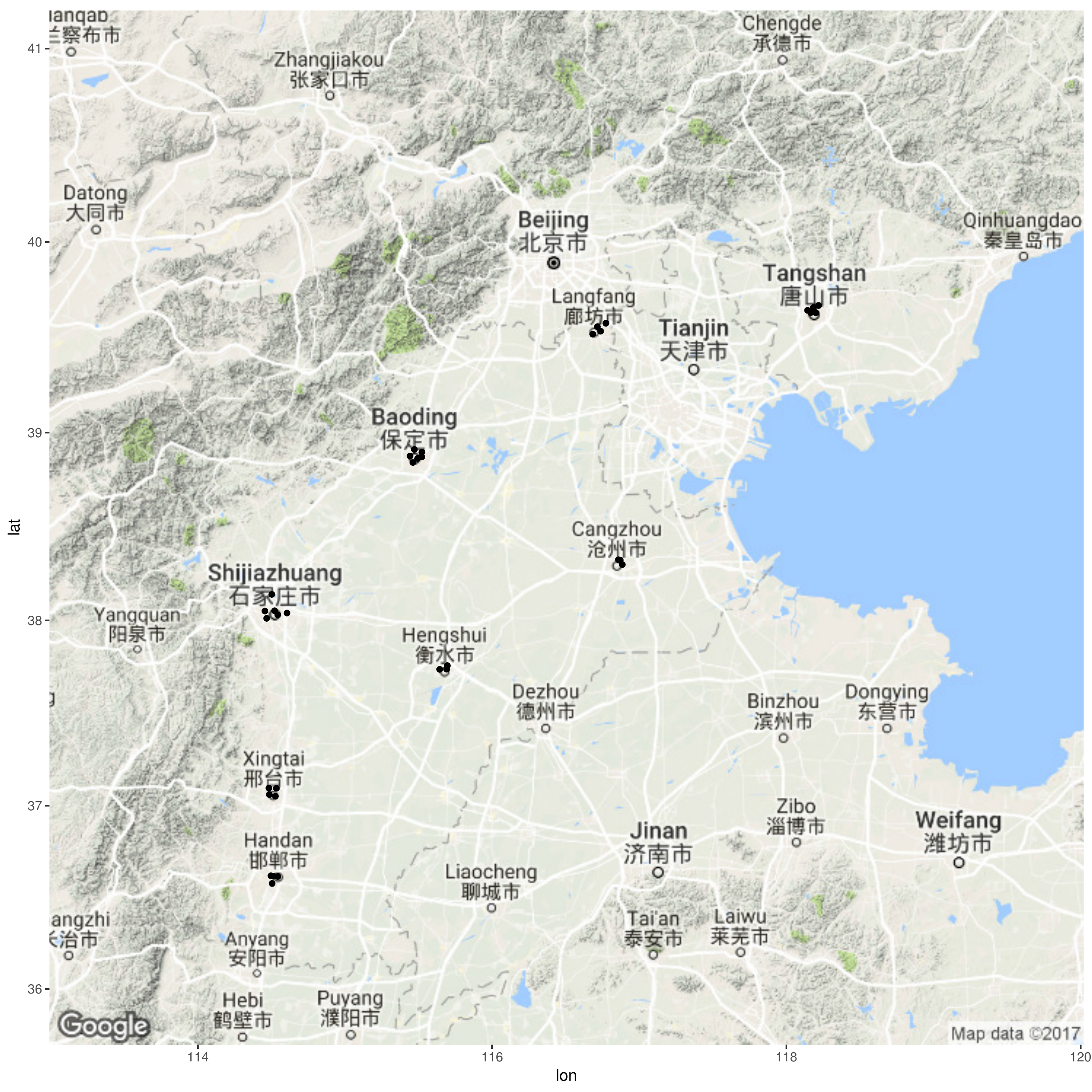}}
	\caption{\label{clustinmap}Spatial locations of the 36 monitoring stations in Hebei, China (black dots).}
\end{figure}

In this study, we collect the hourly concentration of $PM_{2.5}$ (in $\mu g/m^3$) from 36 monitoring stations from 8 cities in Hebei for three seasonal years from March 2014 to February 2017. The spatial locations of the 36 stations are shown in Figure \ref{clustinmap} (highlighted with black dots). 

For covariates, firstly, we consider one-hour lagged concentration of other four pollutant gases: sulfur dioxide ($SO_2$), nitrogen dioxide ($NO_2$), ozone ($O_3$), and carbon monoxide ($CO$), since a great portion of $PM_{2.5}$ contains secondary particles, which are formed in the atmosphere through chemical reactions involving other accumulated gases in the past few hours \citep{wang2013sulfate}. The four gases are all measured in $\mu g/m^3$. Secondly, we collect meteorological data: barometric pressure ($PRES$, in hectopascal), air temperature ($TEMP$, in degree celsius), dew point temperature ($DEWP$, in degree celsius), integrated rainfall ($IRAIN$, in millimeter), integrated wind speed ($Iws$, in meter per second), and wind directions in four categories (northeast, northwest, southeast, and south west), as well as the interaction between integrated wind speed and wind directions. Previous studies have shown that there are non-neglectable meteorological impacts on $PM_{2.5}$ \citep{liang2015assessing}. Moreover, we measure the distance of each site to the Taihang Mountains ($dist\_to\_moun$, in kilometer) to further eliminate the influence of different dispersion conditions caused by topography. 

Both the meteorological variables and pollutants are recorded in the same monitoring network, which can be matched for each monitoring station. 
In Table \ref{SiteinHebei}, we show each station's number, name, and the city where it locates. Due to significantly different patterns of $PM_{2.5}$ across the four seasons, and the association between air pollution and the meteorological variables can also vary across seasons (e.g. \cite{zheng2005seasonal}, \cite{ito2007seasonal} and \cite{liang2015assessing}), we estimate the model for each season respectively. The stations with missing value rate of $PM_{2.5}$ more than 20\% in one season are removed from the data set. All the variables are standardized to ease the comparison of their effects. 

\begin{table}
	\caption{ \label{SiteinHebei} 36 monitoring stations in Hebei Province, displaying in specific city and location}
	\centering
	\begin{tabular}{@{} *{3}{l} @{} }
		\toprule
		{\bf Site}& {\bf City}&{\bf Site Names}\\
		\hline
		1$\sim$6&Baoding&Dibiaoshuichang~~Huadianerqu~~Jiancezhan\\&&Jiaopianchang~~Jiedaizhongxin~~Youyongguan\\
		\hline
		7$\sim$9&Cangzhou&Cangxianchengjianju~~Dianshizhuanbozhan\\&&Shihuanbaoju\\
		\hline
		10$\sim$13&Handan&Congtaigongyuan~~Dongwushuichulichang\\&&Huanbaoju~~Kuangyuan\\
		\hline
		14$\sim$16&Hengshui&Dianjibeichang~~Shihuanbaoju~~Shijiancezhan\\
		\hline
		17$\sim$20&Langfang&Beihuahangtianxueyuan~~Jiancezhongxin\\&&Kaifaqu~~Yaocaigongsi\\
		\hline
		21$\sim$26&Shijiazhuang&Gaoxinqu~~Renminhuitang~~Shijigongyuan\\&&Xibeishuiyuan~~Xinangaojiao~~Zhigongyiyuan\\
		\hline
		27$\sim$32&Tangshan&Gongxiaoshe~~Leidazhan~~Shierzhong\\&&Taocigongsi~~Wuziju~~Xiaoshan\\
		\hline
		33$\sim$36&Xingtai&Dahuoquan~~Luqiaogongsi~~Shihuanbaoju\\&&Xingshigaozhuan\\
		\bottomrule
	\end{tabular}
\end{table}

\section{Methodology}

In this section, we first introduce the setting of the hidden dynamic geostatistical calibration model. In the model, we add calibration components to account for the heterogeneity among readings from different stations. Next we introduce Kalman filter and smoother to obtain the estimation of hidden random variable. Then the main process of EM algorithm \citep{krishnan1997algorithm} is discussed to obtain the maximum likelihood estimates of model parameters. Eventually, computational information matrix is evaluated to assess the accuracy of model parameters.

\subsection{The hidden dynamic geostatistical calibration model}

Denote $y(s,t)$ as the measured concentration of $PM_{2.5}$ at site $s \in \mathcal{D}$, and a discrete time $t \in 1,...,T$. Denote $\mathcal{D}$ as the set of locations of an environmental monitoring network with $n$ sties, each of which is identified by its latitude and longitude coordinates. Exogenous variables, or fixed impact can enter the observation equation (\ref{eqn:Eq11}) to make it more extensive. Hence, we include fixed effects $\mathbf{X}(s, t)$ of $k$ dimension to account for the effects of other pollutants, meteorological variables and geographic information. The residual $\varepsilon (s , t )$ is a time-space independent Gaussian measurement error with variance $\sigma^2$. We also assume that $\varepsilon$ is independent of $z$ and $\eta$. Our model can be expressed as follows,

\begin{eqnarray}
\label{eqn:Eq11}
y(s, t) &=& \mathbf{X}(s,t)\mathbf{\beta } + \alpha_s z(s,t) + \varepsilon (s,t),\\
\label{eqn:Eq12}
z(s,t) &=& g z(s, t-1 ) + \eta (s,t).
\end{eqnarray}

In this model, $y(s, t)$ represents the observed $PM_{2.5}$ concentration at location $s$ and time $t$. After ruling out the influences from other pollutants, meteorological factors and dispersion conditions, the random effect $z( s, t )$ can be considered as the ``true'' emission of $PM_{2.5}$ at location $s$ and time $t$ as discussed in \cite{Fass2007Air}. The $\alpha_s$ are the calibration components accounting for heterogeneity among readings of different stations. The relationship between the ``true'' emission and observed $PM_{2.5}$ is assumed to be similar in trend for all the stations. That is, as long as there is no instrumental error or interference, the calibration component $\alpha_s$ should be similar for all stations $s$. 

The state equation (\ref{eqn:Eq12}) determines the rule for generation of latent variable $z(s,t)$ from past $z(s,t-1)$, reflected by an order one autogression $g$. The innovation $\eta (s , t )$ is a sequence of unit-variance time homogenous Gaussian random fields, with exponential spatial covariance function $\rho(\theta) = exp( (-||s - s^{\prime }|| / \theta) $, where $||s - s^{\prime } || $ is the Euclidean distance between $s$ and $s^{\prime} \in \mathcal{D}$, and $\theta$ is a scale parameter. 

To give the matrix presentation, we introduce the observations at time $t$ stored in the $n$ dimension vector
\[ y_t = (y(s_1,t), ..., y(s_n,t))'. \]
Hence we can rewrite the model in matrix notation as follows
\begin{eqnarray}
\nonumber &&y_t = \mu_{t} + \epsilon_t,\\
&&\mu_{t}=\mathbf{X}_t\beta + A z_t,\\
\nonumber&& z_t = g z_{t-1} + \eta_t
\end{eqnarray}
where $\mathbf{X}_t = (\mathbf{X}(s_1,t)',...,\mathbf{X}(s_n,t)')'$ is a design matrix of dimension $n \times k$, with elements $\mathbf{X}(s_i,t)'$ given by $k\time 1$ dimension vector in analogy to equation \ref{eqn:Eq11}. Moreover, the random effect $z_t = (z(s_1,t), ..., z(s_n,t))'$ is a also $n$ dimension vector.
Correspondingly, the calibration parameter $A$ is an $n \times n$ matrix given by 
$$A = diag(\alpha_1,...,\alpha_{n})$$

Based on the matrix presentation, we show how to obtain unobserved latent variable, parameter estimation and variance covariance matrix. 
Since the HDGC model contains random effect and air pollution data has missing values, the expectation maximization (EM) algorithm is a natural choice for parameter estimation \citep{smith2003spatiotemporal}. Consistency of the estimators, especially the $\alpha_s$'s, are guaranteed under Gaussianity assumptions of the random effects (see Chapter 6 in \cite{sumway2006time}; \cite{wu1988strong}). 

\subsection{Kalman filter and smoother}

We define the conditional mean $z_t^s$ and conditional variance covariance $P_{t_1, t_2}^s$ for the underlying hidden variable $z_t$, given the observed data $Y_s = \{ y_1, ..., y_s\}$ to time $s$,
\begin{eqnarray}
\nonumber &&z_t^s = E(z_t|y_1,...,y_s),\\
\nonumber &&P_{t_1, t_2}^s = E\{(z_{t_1} - z_{t_1}^s) (z_{t_2} - z_{t_2}^s)'\}. 
\end{eqnarray}
For simplicity, we use the notation $P_{t}^s$ when $t_1=t_2$. 

The Kalman filter specifies how to update from $z_{t-1}^{t-1}$ to $z_{t}^{t}$ once the new observation $y_t$ is obtained, without having to reprocess the entire dataset $\{ y_1,.., y_t\}$. The filtered values are 
\begin{eqnarray}
\nonumber &&z_t^t = z_t^{t-1} + K_t (y_t - \mathbf{X}_t \beta - A z_{t}^{t-1}),\\
\nonumber &&P_t^t = P_t^{t-1} - K_t A P_t^{t-1},
\end{eqnarray}
where $K_t = P_t^{t-1} A (A P_t^{t-1} A' + \Sigma_{\epsilon})^{-1}$ is called Kalman gain. 
It is obvious the Kalman predicted value should be given by 
\begin{eqnarray}
\nonumber &&z_t^{t-1} = g z_{t-1}^{t-1},\\
\nonumber && p_t^{t-1} = g^2 P_{t-1}^{t-1} + \Sigma_{\eta}.
\end{eqnarray}
Now we consider the problem of obtaining estimators for $z_t$ based on the entire data sample $\{y_1,..., y_T\}$, namely $z_{t}^{T}$. The Kalman smoother recursion for $t = T,...,1$ are as follows,
\begin{eqnarray}
\nonumber &&z_{t-1}^{T} = z_{t-1}^{t-1} + g P_{t-1}^{t-1}  (P_t^{t-1})^{-1} (z_t^{T} - z_t^{t-1}),\\
\nonumber &&P_{t-1}^{T} = P_{t-1}^{t-1} + g^2 P_{t-1}^{t-1} (P_t^{t-1})^{-1} (P_t^{T} - P_t^{t-1}) (P_t^{t-1})^{-1} P_{t-1}^{t-1}
\end{eqnarray}

\subsection{EM algorithm}

Suppose for observed data $Y=\{y_{1},...,y_{T}\}$ and latent $Z=\{z_{0},z_{1},...,z_{T}\}$, we have the following conditional Gaussian probability distributions,

\begin{description}
	\centering
	\item $y_{t}|z_{t}\sim N_{n}(\mu _{t},\Sigma _{\varepsilon }),$
	\item $z_{t}|z_{t-1}\sim N_{n}(g z_{t-1},\Sigma _{\eta }),$
	\item $z_{0}\sim N_{n}(\mu _{0},\Sigma _{0}).$
\end{description}

Let the unknown parameter set be $\Psi  = ( \beta, \alpha, \sigma^2, g, \theta )$. The complete-data log-likelihood function $l(\Psi ;Y,Z)$ can be written as,
\begin{eqnarray*}
	-2l(\Psi ;Y,Z) &=&\log |\Sigma
	_{0}|+\sum_{t=1}^{T}(z_{0}-\mu _{0})^{\prime }\Sigma _{0}^{-1}(z_{0}-\mu
	_{0}) \\
	&&+T\log |\Sigma _{\varepsilon }|+\sum_{t=1}^{T}(y_{t}-\mu _{t})^{\prime
	}\Sigma _{\varepsilon }^{-1}(y_{t}-\mu _{t})\\
	&&+T\log |\Sigma _{\eta }|+\sum_{t=1}^{T}(z_{t}- g z_{t-1})^{\prime }\Sigma
	_{\eta }^{-1}(z_{t}- g z_{t-1}).
\end{eqnarray*}

The initial $\Psi_0$ are set as $\beta^{<0>}$ estimated from least square estimate, $\alpha_s^{<0>} = 0.5$ for all stations $s = 1,...,n$, and the other parameters $(\sigma^{2})^{<0>} = 0.1, g^{<0>}=0.2, \theta^{<0>}=100$. 
Conditioning on the observed data $Y = \{y_1, \dots, y_T\}$, we calculate the expectation of the complete data log-likelihood function under the parameter $\Psi ^{<m>}$,
\begin{eqnarray*}
	Q(\Psi ,\Psi ^{<m>}) &=&E_{\Psi ^{<m>}}[-2l(\Psi ;Y,Z)|Y].
\end{eqnarray*} 

At the M-step, the parameter set $\Psi ^{<m+1>}$ maximizes the conditional expectation function. It can be shown that, for the $m+1$ iteration, the parameter set is updated as below,

\begin{description}
	
	\item
	\begin{equation}
	\alpha_{s}^{<m+1>}=\frac{\sum_{t=1}^{T}tr((diag(e_{s})z_{t}^{T})(y_{t}-\mathbf{X}_t \beta^{<m>} )^{\prime })}{
		\sum_{t=1}^{T}tr((z_{t}^{T}(z_{t}^{T})^{\prime}+P_{t}^{T})diag(e_{s}))},
	\end{equation}
	where $e_s$ is the unit vector with the s-th component equals to 1.
	
	\item
	\begin{equation}
	\beta^{<m+1>} =[\sum_{t=1}^{T}(\mathbf{X}_t)^{\prime }\Sigma _{\varepsilon}^{-1,<m>}\mathbf{X}_t]^{-1}[\sum_{t=1}^{T}(\mathbf{X}_t)^{\prime }\Sigma _{\varepsilon}^{-1,<m>}(y_{t}-A^{<m>}z_{t}^{T})],
	\end{equation}
	
	\item
	\begin{equation}
	(\sigma^{2})^{<m+1>}=\frac{1}{nT} \sum_{t=1}^{T}
	tr(\Omega _{t}^{<m>}),
	\end{equation}
	
	\item
	\begin{equation}
	g^{<m+1>}=\frac{tr(\Sigma _{\eta }^{-1}S_{10})}{tr(\Sigma _{\eta }^{-1}S_{00})},
	\end{equation}
	
	\item
	\begin{equation}
	\theta^{<m+1>} =  \arg_{max} ~~ T\log |\Sigma _{\eta }|+tr[\Sigma _{\eta }^{-1}(S_{11}-S_{10}g - g S_{10}^{\prime } + g^2 S_{00} )].
	\end{equation}
\end{description}
where 
\begin{eqnarray}
\nonumber && \Omega_0 = z_{0}^T (z_{0}^T)^{\prime }+P_0^T,
~\Omega_t = (y_{t}-\mathbf{X}_t \beta- A z_{t}^{T}) (y_{t}-\mathbf{X}_t \beta
- A z_{t}^{T})^{\prime} + A P_{t}^{T} A^{\prime },\\
\nonumber && S_{10} = \sum_{t=1}^{T}z_t^{T}(z_{t-1}^{T})^{\prime }+ P_{t,t-1}^{T},
~ S_{00} = \sum_{t=1}^{T} z_{t-1}^{T}(z_{t-1}^{T})^{\prime }+ P_{t-1}^{T},\\
\nonumber && S_{11} = \sum_{t=1}^{T} z_t^T(z_t^T)^{\prime }+ P_{t}^{T}. 
\end{eqnarray}

Since the geographic parameter $\theta$ can not be solved in a closed form, we update $\theta$ through numerical optimization by the Nelder-Mead algorithm. The EM algorithm stops when the change of Euclidean distance of the parameter set and log likelihood are both significantly small. 
\subsection{Computational information matrix from marginal likelihood}

The direct likelihood computation is based on $nT\times nT$ matrix, which is exceedingly high dimension. Therefore, based on the kalman filter we calculate the information matirx, which is computationally affordable as it requires $T$ inversions of $n \times n$ dimensional matrices. Standard results in \cite{sumway2006time} 6.3 allow the computational information matrix. And we define 
\begin{eqnarray}
\nonumber &&\tilde{\epsilon_t} = \tilde{\epsilon_t}(\Phi) = y_t - X_t \beta - A z_{t}^{t-1}\\
\nonumber &&\Sigma_{t} = \Sigma_{t}(\Phi) = A P_{t-1}^{t} A' + \Sigma_{\epsilon}
\end{eqnarray}
where $z_{t}^{t-1} = z_{t}^{t-1}(\Phi) $ and $P_{t}^{t-1} = P_{t}^{t-1} (\Phi)$
are the Kalman filter outputs at convergence. 
The loglikelihood function excluding an additive constant is
$$ -2log(L(\Phi)) = \sum_{t=1}^{T} log|\Sigma_{t}| + \sum_{t=1}^{T} \tilde{\epsilon_t}' \Sigma_{t}^{-1} \tilde{\epsilon_t} $$

Denote the information matrix $\hat{\mathcal{J}}$, we can obtain the variance-covariance matrix by,
$$ V(\hat{\Phi}) \cong \hat{\mathcal{J}}^{-1} $$
where 
\begin{eqnarray*}
	\hat{\mathcal{J}}_{i,j} & \cong & \sum_{t=1}^{T}\{ (\partial_i \tilde{\epsilon_t}') \Sigma_{t}^{-1} (\partial_j \tilde{\epsilon_t}) + \\
	& + & \frac{1}{2} tr\{\Sigma_{t}^{-1} (\partial_i \Sigma_{t}) \Sigma_{t}^{-1} (\partial_j \Sigma_{t}) \} \\
	& + & \frac{1}{4} tr\{ \Sigma_{t}^{-1} (\partial_i \Sigma_{t})\} tr\{ \Sigma_{t}^{-1} (\partial_j \Sigma_{t})\} \}
\end{eqnarray*}
and the involved derivatives are given in the sequel given by recursions.
\begin{itemize}
	
	\item $\partial_i \tilde{\epsilon_t} = - X_t^{\beta} \partial_i\beta - A \partial_i z_t^{t-1} - (\partial_i \alpha) z_t^{t-1}$,
	
	\item $\partial_i z_t^{t-1} = (\partial_i G) z_{t-1}^{t-2} + G \partial_i z_{t-1}^{t-2}
	+ (\partial_i K_{t-1}) \tilde{\epsilon_t} + K_{t-1} \partial_i \tilde{\epsilon_t}$,
	
	\item $\partial_i \Sigma_t = (\partial_i \alpha)P_{t}^{t-1}A + A(\partial P_{t}^{t-1})A + AP_{t}^{t-1}(\partial_i \alpha) + \partial_i \Sigma_{\epsilon}$,
	
	\item $\partial_i K_t = \left[ (\partial_i G) P_t^{t-1} A + G(\partial_i P_t^{t-1})A + GP_t^{t-1}(\partial_i \alpha) - K_t(\partial_i\Sigma_{t}) \right] \Sigma_{t}^{-1}$,
	
	\item $\partial_i P_{t}^{t-1} = (\partial_i G) P_{t-1}^{t-2}G' + G( \partial_i P_{t-1}^{t-2}) G' + G(P_{t-1}^{t-2})(\partial_i G') + \partial_i\Sigma_{\eta} \\~~~~~~~~~~~~~~~ - (\partial_i K_{t-1}) \Sigma_{t} K_{t-1} - K_{t-1} (\partial_i \Sigma_t) K_{t-1}' - K_{t-1} \Sigma_{t} (\partial_i K_{t-1}')$,
	
\end{itemize}

The closed form for the derivativers is given in the following to compute the above recursions,
\begin{itemize}
	\item $\partial_i \beta_j=1$ if $\Phi_i = \beta_j$ and $\partial_i \beta_j=0$ else;
	
	\item $\partial_i \alpha_j=1$ if $\Phi_i = \alpha_j$ and $\partial_i \alpha_j=0$ else;
	
	\item $\partial_i g =1$ if $\Phi_i = g$ and $\partial_i G =0$ else;
	
	\item $\partial_i \Sigma_{\eta} =1$ if $\Phi_i = \theta$ and $\partial_i \Sigma_{\eta } = 0$ else;
	
	\item $\partial_i \Sigma_{\epsilon} =1$ if $\Phi_i = \sigma^2$ and $\partial_i \Sigma_{\epsilon} =0$ else;
	
\end{itemize}

\section{Simulation study}

\begin{figure}	
	\centering
	\includegraphics[width=5cm]{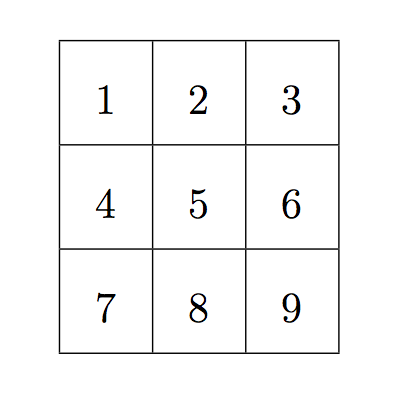}
	\caption{ \label{gridcell} Grid cell of 9 stations.}
\end{figure}

To demonstrate the ability of the calibration component $\alpha_s$ to detect biased readings from certain site, we conduct the following simulation studies. The responses  ${y(s,t)}$ are generated from the model with a $3\times 3$ grid cell, 

\begin{eqnarray}
\label{Eq2}
y(s, t) = \alpha_s z(s,t) + \varepsilon (s,t),\\
\label{Eq3}
z(s,t) = g z(s, t-1 ) + \eta (s,t),
\end{eqnarray}
where $g=0.5$, $s=1,2,...,9$ and $t=1,2,...,100$. The residuals $\varepsilon (s , t )$ in equation (\ref{Eq2}) are independently generated from a common Gaussian distribution with zero mean and $\sigma=0.1$. The innovation ${\eta (s , t )}$ in equation (\ref{Eq3}) is generated from a unit-variance time homogenous Gaussian random fields, with scale parameter $\theta=100$ used for the exponential spatial covariance. The 9 stations are arranged in Figure \ref{gridcell}. 

We consider two different scenarios. In the first scenario, we assume site 5 in the center is the site with biased readings. We set $\alpha_5=0.5$, while $\alpha=0.8$ for other stations. In the second scenario, the site with biased readings is at the corner (site 3), with $\alpha_3=0.5$, while $\alpha=0.8$ for other stations. For each scenario, we report the empirical means, standard error, and 95\% confidence intervals (CIs) of the estimated parameters in Table \ref{SR1} and \ref{SR2} based on 1000 simulation. From Table \ref{SR1} and \ref{SR2}, we can see that no matter where the suspicious site locates, it can be identified by the estimated $\alpha$. Moreover, other parameters $g$, $\theta$ and $\sigma$ can be accurately estimated. 

\begin{table} 
	\caption{ \label{SR1} Scenario 1. Site 5 is the site with biased reports. Based on 1000 simulation, we show mean, standard deviation and 95\% lower and upper bounds of estimate results. }
	\centering
	\resizebox{\textwidth}{!}{\begin{tabular}{ccccccccccccc}
			\hline 
			&$\alpha_1$&$\alpha_2$&$\alpha_3$&$\alpha_4$&$\alpha_5$&$\alpha_6$&
			$\alpha_7$&$\alpha_8$&$\alpha_9$&g&$\theta$&$\sigma$\\
			\hline
			T=100&0.8&0.8&0.8&0.8&0.5&0.8&
			0.8&0.8&0.8&0.5&100&0.1\\
			\hline
			Mean&0.828&0.828&0.826& 0.830& 0.522& 0.831&0.828&0.828& 0.830 &0.495&95&0.108\\
			Sd&0.048 &0.049&0.049&0.048&0.046&0.050&0.048 &0.049&0.048 &0.041&10& 0.005\\
			LB&0.734 &0.734 &0.726&0.735 &0.433 &0.723&0.731& 0.734 &0.718& 0.411&77& 0.098\\
			UB&0.928 &0.927&0.923&0.924&0.612&0.933& 0.929 &0.930&0.926 &0.571&117&0.121\\
			\hline
			T=500&0.8&0.8&0.8&0.8&0.5&0.8&
			0.8&0.8&0.8&0.5&100&0.1\\
			\hline
			Mean&0.825&0.824& 0.825& 0.824&0.521&0.825&0.825& 0.823 & 0.825& 0.500& 95&0.106\\
			Sd& 0.019&0.018& 0.019&0.018&0.018& 0.018&0.018 & 0.019 & 0.019 & 0.017&4.4&0.002\\
			LB&0.788&0.790& 0.790 &0.786&0.485&0.789 & 0.792& 0.786& 0.789&0.465 &86& 0.102\\
			UB&0.866&0.862& 0.867&0.860&0.559&0.861& 0.864& 0.865 & 0.864 &0.534&103& 0.110\\
			\hline
	\end{tabular}}
\end{table}
\begin{table}
	\caption{\label{SR2} Scenario 2. Site 3 is the site with biased reports. Based on 1000 simulation, we show mean, standard deviation and 95\% lower and upper bounds of estimate results.}
	\centering
	\resizebox{\textwidth}{!}{\begin{tabular}{ccccccccccccc}
			\hline 
			&$\alpha_1$&$\alpha_2$&$\alpha_3$&$\alpha_4$&$\alpha_5$&$\alpha_6$&
			$\alpha_7$&$\alpha_8$&$\alpha_9$&g&$\theta$&$\sigma$\\
			\hline
			T=100&0.8&0.8&0.5&0.8&0.8&0.8&
			0.8&0.8&0.8&0.5&100&0.1\\
			\hline
			Mean&0.831& 0.827& 0.524& 0.826& 0.827& 0.828&  0.826& 0.826&  0.828&  0.497& 95& 0.107\\
			Sd&0.046& 0.048&  0.048& 0.048&   0.047&  0.047& 0.047& 0.049&  0.048& 0.041& 10&  0.005\\
			LB&0.739& 0.730&  0.429& 0.724&  0.727& 0.732&  0.728& 0.728&  0.727& 0.414&  78&  0.097\\
			UB&0.925& 0.921&  0.623&  0.919& 0.925& 0.918& 0.921& 0.926& 0.925& 0.571&115&  0.117\\
			\hline
			T=500&0.8&0.8&0.5&0.8&0.8&0.8&
			0.8&0.8&0.8&0.5&100&0.1\\
			\hline
			Mean&0.825& 0.824&  0.524&0.823&   0.821&0.823&  0.824& 0.822&   0.825&  0.500& 94&  0.106\\
			Sd& 0.018& 0.017& 0.019&  0.018& 0.017&  0.017&  0.018& 0.018&  0.018&  0.017&  4.5& 0.002\\
			LB&0.792& 0.791&0.486&  0.788& 0.787& 0.790&  0.788& 0.788&0.790&   0.463&  86&    0.102\\
			UB&0.864&    0.861&   0.563&    0.862&    0.856&    0.862&    0.862&    0.860&   0.863&    0.537&  103&    0.110\\
			\hline
	\end{tabular}}
\end{table}

The simulations results illustrate that the HDGC model is able to reveal those stations whose readings have different patterns from their neighbors. Although we have spatially-correlated time series data, the EM algorithm can still provide consistent estimates of model parameters (e.g. \cite{smith2003spatiotemporal}; \cite{fasso2009algorithm}).

\section{Analysis of $PM_{2.5}$ pollution in Hebei}

In the following sections, we show and interpret the parameter estimates as well as the goodness of fit of the model. We then compare the estimated $\alpha$ values for all stations (see Table \ref{alphanew1}), and check for the stations with consistent outlying values. Note that lower $\alpha$ values indicate that observed PM2.5 substantially deviates from true $PM_{2.5}$, which calls for further investigation of these stations. 

\subsection{Covariates impact}

Table \ref{betanew} shows the estimated values for parameters $\beta, g, \theta, \sigma^2$ at each season over three years. Since all the variables in our model are standardized, the estimated $\beta$ for each covariate can provide insight in its contribution to $PM_{2.5}$ level. Most of the $\beta$ estimates are statistical significant in each season, indicating the complicated formation mechanism of $PM_{2.5}$ pollution. 

The root of mean square error of model residuals ranges from $3.69$ to $15.14$, with a mean of $8.69$, which is shown in the last second line in Table \ref{betanew}. 
Comparing to the mean level of the measured $PM_{2.5}$ of the 12 seasons, which ranges from $52$ to $158$ with an overall mean of $94$, we claim that our model has a decent goodness of fit. The last line of the table shows missing rate of $PM_{2.5}$ for all 36 monitoring stations, with a maximum of $0.056$ in autumn 2016.

\begin{table}
	\caption{ \label{betanew} Parameter $\beta, g ,\theta, \sigma^2 $ estimates and their levels of significance (* for p-value $ <0.05$; ** for p-value $ <0.01$; *** for p-value $<0.001$) from spring 2014 to winter 2016. }
	\centering
	\resizebox{\textwidth}{!}{
		\begin{tabular}{@{} cccccccccccccc @{}} \hline 
			\multirow{2}{*}{Period} &
			\multicolumn{4}{c}{Year 2014} &
			\multicolumn{4}{c}{Year 2015} &
			\multicolumn{4}{c}{Year 2016} \\
			\cline{2-13}
			& Spring& Summer & Autumn& Winter
			& Spring& Summer & Autumn& Winter
			& Spring& Summer & Autumn& Winter\\
			\hline
			$SO_2$&0.05***&0.02***&0.02***&0.08***&
			0.06***&0.03***&0.10***&0.00&
			0.07***&0.00&0.05***&0.02***\\
			$NO_2$&0.15***&0.09***&0.19***&
			0.20***&0.13***&0.09***&0.26***&
			0.18***&0.21***&0.06***&0.22***&0.22***\\
			$O_3$&-0.05***&-0.05***&-0.01***&0.01***&
			-0.05***&0.02***&0.18***&0.03***&
			0.12***&0.06***&0.11***&0.01***\\
			$CO$&0.07***&0.06***&0.08***&0.26***&
			0.09***&0.07***&0.29***&0.17***&
			0.10***&0.03***&0.19***&0.28***\\
			$PRES$&0.03&-0.13***&-0.11***&-0.06***&
			0.03&-0.08***&0.01&-0.09***&
			-0.17***&-0.01&-0.12***&-0.10***\\
			$TEMP$&-0.22***&-0.00&-0.33***&-0.09***&
			-0.17***&0.06***&-0.33***&-0.17***&
			-0.47***&-0.11***&-0.51***&-0.12***\\
			$DEWP$&0.32***&0.32***&0.29***&0.28***&
			0.33***&0.34***&0.21***&0.39***&
			0.37***&0.34***&0.34***&0.24***\\
			$IRAIN$&-0.04***&-0.00&-0.04***&-0.01***&
			-0.03***&-0.01***&-0.05***&-0.03***&
			-0.06***&-0.05***&-0.07***&-0.03***\\
			$Iws$&0.01&0.01&0.05**&0.01&
			0.03&0.05&-0.03&-0.01&
			0.13***&-0.01&-0.06***&0.07***\\
			$NE$&0.01&0.01&0.02***&0.01***&
			0.01&0.01&0.02**&0.01***&
			0.03***&0.02***&0.02***&0.03***\\
			$NW$&0.02***&0.01&0.02***&0.00&
			0.03***&0.01&-0.01&0.01**&
			0.04***&0.01*&0.03***&0.02***\\
			$SE$&0.00&-0.01&0.03***&0.02***&
			-0.00&-0.00&0.04***&0.02***&
			0.05***&-0.01&0.02***&0.02***\\
			$SW$&0.01&0.01**&0.00&0.00&
			0.01&-0.00&-0.01*&0.01&
			0.01&-0.00&-0.01**&0.00\\
			$Iws*NE$&-0.01&-0.01&-0.05***&-0.03*&
			-0.04&-0.07&-0.09&-0.05**&
			-0.13***&-0.03**&-0.02&-0.11***\\
			$Iws*NW$&-0.04&-0.01&-0.07***&-0.01&
			-0.08*&-0.07&-0.01&-0.04&
			-0.16***&-0.01&-0.03**&-0.10***\\
			$Iws*SE$&-0.02&0.00&-0.08***&-0.07***&
			-0.02&-0.05&-0.04&-0.04**&
			-0.16***&0.02*&0.00&-0.09***\\
			$Iws*SW$&-0.02&-0.02&-0.03*&-0.01&
			-0.04&-0.01&0.05&-0.02&
			-0.12***&0.01&0.07***&-0.06***\\
			$Dist\_to\_Moun$&-0.08**&-0.02&-0.09**&-0.17***&
			-0.01&-0.10***&-0.08**&-0.07***&
			-0.03&0.01&-0.12***&-0.14***\\
			g&0.94***&0.92***&0.96***&0.92***&
			0.94***&0.94***&0.94***&0.93***&
			0.93***&0.93***&0.94***&0.93***\\
			$\theta$ (km)&17***&13***&25***&17***&
			52***&26***&111***&23***&
			110***&9***&125***&160***\\
			$\sigma^2$&0.03&0.05&0.01&0.02&
			0.05&0.04&0.03&0.02&
			0.05&0.03&0.03&0.02\\
			\bf root of MSE & 8.94 & 7.79 & 7.34 & 8.27
			& 9.28 & 5.49 & 10.38 & 9.40
			&9.14 & 3.69 & 9.47 & 15.14\\
			\bf Missing Rate & 0.019 & 0.026 & 0.032 & 0.017 &
			0.011 & 0.021 & 0.031 & 0.010 &
			0.013 & 0.043 & 0.056 & 0.054\\
			\hline
	\end{tabular}}
\end{table}

The three gaseous precursors $SO_2$, $NO_2$, and $CO$ almost all have significantly positive effects (with $p$-values less than 0.001) on $PM_{2.5}$ pollution. 
Among them, $NO_2$ has the largest influence on average. 
The second large effect on observed $PM_{2.5}$ is from $CO$, and $CO$ has the similar impact as $NO_2$ in autumn and winter. 
Besides, we cannot determine the exact effect of $O_3$ on $PM_{2.5}$. $O_3$ is predominantly produced by photochemical reactions involving two major precursors, volatile organic compounds ($VOC$) and oxides of nitrogen ($NO_x$). An inverse relationship has been found between $O_3$ and $NO_2$ (e.g. \cite{de2010air}, \cite{han2011analysis} and \cite{sillman1999relation}). Hence, a positive association between $PM_{2.5}$ and $NO_2$ may cover the relationship between $PM_{2.5}$ and $O_3$.

For meteorological variables, dew point has the strongest positive effect on $PM_{2.5}$ among all meteorological variables. An increase in dew point temperature indicates the amount of water vapor in the atmosphere is increasing, which would promote the formation of $PM_{2.5}$ \citep{yang2011characteristics}. Mostly, temperature is negatively related with $PM_{2.5}$.
Rainfall can help dilute the haze and increase the sedimentation of aerosols, and thus significantly reduce $PM_{2.5}$ concentration \citep{jacob2009effect}. The barometric pressure is found to have a negative effect on $PM_{2.5}$, although not always significant, which is similar to the findings in \cite{liang2015assessing}. Integrated wind speed interacted with directions can disperse and dilute $PM_{2.5}$, especially in the winter heating period \citep{dawson2007sensitivity}. 

As expected, the distance to Taihang Mountain ($dist\_to\_moun$) mostly has a significant negative effect on $PM_{2.5}$, since it can reflect the dispersion condition caused by topography. That is the nearer to Taihang Mountain, the easier to accumulate the polluted air, while the farther, the easier to disperse. 

In addition, the mode number of estimated parameter $g$ is $0.94$, which is the lag one coefficient of random effect $z(s,t)$, and thus reveals a strong temporal persistency of hourly air pollution data. Range parameter $\theta$ presents the spatial correlation, and from Table \ref{betanew} we can conclude that the spatial stations are correlated in city range (around $20$ kilometers) in most case, whereas have become more correlated since the end of 2016.  

\subsection{Comparison of $\alpha$ estimates}

In this section, we specially analyze the estimated values of $\alpha$, which indicate the relationships between the observed and the ``true'' PM2.5 level for each site. The estimated $\alpha$ values and their standard deviation of the 36 monitoring stations for 12 seasons are shown in Table \ref{alphanew1} and Table \ref{alphanew2}. 

\begin{table}
	\tiny
	\caption{ \label{alphanew1} $\mathbf{\alpha}$ estimates and standard deviation in parentheses for twelve seasons from spring 2014 to winter 2016.} 
	\resizebox{\textwidth}{!}{
		\begin{tabular}{lclclclclclclclclclclclcl} 
			\toprule
			\multirow{2}{*}{Site} &
			\multicolumn{4}{c}{Year 2014} &
			\multicolumn{4}{c}{Year 2015} &
			\multicolumn{4}{c}{Year 2016} \\
			\cline{2-13}
			& Spring& Summer & Autumn& Winter
			& Spring& Summer & Autumn& Winter
			& Spring& Summer & Autumn& Winter\\
			\toprule
			1&0.5860&0.6817&0.6192&0.5687&0.4239&0.3158&0.4566&0.4522&0.3616&0.2852&0.4682&0.4229\\
			&(0.0153)&(0.0151)&(0.0134)&(0.0137)&(0.0174)&(0.0187)&(0.0149)&(0.0138)&(0.0177)&(0.0205)&(0.0147)&(0.0146)\\
			\hline
			
			2&0.4002&0.4539&0.4609&0.5896&0.3988&0.3452&0.5588&0.6891&0.5423&0.6132&0.4628&0.4933\\
			&(0.0171)&(0.0173)&(0.0134)&(0.0139)&(0.0175)&(0.0176)&(0.0141)&(0.0127)&(0.0151)&(0.0145)&(0.0146)&(0.0141)\\
			\hline
			
			3&0.5671&0.4082&0.5004&0.4807&0.3505&0.4699&0.4735&0.4434&0.3978&0.7120&0.4544&0.4382\\
			&(0.0150)&(0.0179)&(0.0130)&(0.0141)&(0.0181)&(0.0158)&(0.0145)&(0.0136)&(0.0166)&(0.0138)&(0.0144)&(0.0142)\\
			\hline
			
			4&0.4116&0.4176&1.1242&0.4452&0.3361&0.3086&0.4271&0.4587&0.4087&0.2421&0.3900&0.3979\\
			&(0.0169)&(0.0181)&(0.0122)&(0.0147)&(0.0187)&(0.0189)&(0.0153)&(0.0136)&(0.0167)&(0.0220)&(0.0153)&(0.0149)\\
			\hline
			
			5&0.4343&0.9967&0.5612&0.7375&0.3632&0.2940&0.4890&0.4165&0.4879&0.3020&0.4332&0.4356\\
			&(0.0168)&(0.0138)&(0.0132)&(0.0140)&(0.0185)&(0.0193)&(0.0147)&(0.0142)&(0.0158)&(0.0199)&(0.0149)&(0.0145)\\
			\hline
			
			6&0.9225&1.8331&1.0667&0.4917&0.3870&0.4468&0.5146&0.3844&0.3908&0.3138&0.4672&0.4314\\
			&(0.0134)&(0.0123)&(0.0116)&(0.0139)&(0.0171)&(0.0154)&(0.0141)&(0.0138)&(0.0165)&(0.0184)&(0.0142)&(0.0142)\\
			\hline
			
			7&0.2452&0.2712&0.2071&0.1561&0.2796&0.2778&0.3421&0.2201&0.3077&0.1946&0.3088&0.2526\\
			&(0.0222)&(0.0232)&(0.0197)&(0.0230)&(0.0225)&(0.0205)&(0.0182)&(0.0191)&(0.0203)&(0.0248)&(0.0178)&(0.0185)\\
			\hline
			
			8&0.2335&0.2532&0.2126&0.1384&0.2095&0.3791&0.3758&0.2185&0.3363&0.2165&0.3690&0.2622\\
			&(0.0215)&(0.0223)&(0.0182)&(0.0231)&(0.0240)&(0.0174)&(0.0173)&(0.0184)&(0.0192)&(0.0225)&(0.0165)&(0.0180)\\
			\hline
			
			9&0.2347&0.2900&0.2025&0.1351&0.1977&0.2843&0.3003&0.1743&0.2925&0.1801&0.3413&0.2662\\
			&(0.0213)&(0.0210)&(0.0181)&(0.0228)&(0.0238)&(0.0187)&(0.0184)&(0.0191)&(0.0197)&(0.0233)&(0.0166)&(0.0177)\\
			\hline
			
			10&0.4230&0.4247&1.2155&0.2967&0.3131&0.3978&0.5074&0.5382&0.4179&0.2455&0.4604&0.4495\\
			&(0.0164)&(0.0174)&(0.0117)&(0.0162)&(0.0191)&(0.0164)&(0.0144)&(0.0137)&(0.0165)&(0.0209)&(0.0148)&(0.0144)\\
			\hline
			
			11&0.3060&0.4073&0.3195&0.2597&0.3086&0.3701&0.4769&0.6549&0.4866&0.2722&0.4377&0.4404\\
			&(0.0194)&(0.0184)&(0.0153)&(0.0173)&(0.0199)&(0.0176)&(0.0148)&(0.0135)&(0.0160)&(0.0207)&(0.0151)&(0.0146)\\
			\hline
			
			12&0.3220&0.4212&0.4644&0.4031&0.3901&0.4439&0.4037&0.5642&0.4302&0.2604&0.4298&0.4507\\
			&(0.0177)&(0.0169)&(0.0129)&(0.0141)&(0.0176)&(0.0156)&(0.0153)&(0.0134)&(0.0160)&(0.0196)&(0.0148)&(0.0142)\\
			\hline
			
			13&0.2614&0.3361&0.3934&0.3109&0.3606&0.3852&0.3939&0.5572&0.4860&0.3091&0.4289&0.4377\\
			&(0.0216)&(0.0209)&(0.0146)&(0.0164)&(0.0191)&(0.0176)&(0.0162)&(0.0140)&(0.0163)&(0.0199)&(0.0156)&(0.0149)\\
			\hline
			
			14&0.2391&0.2488&0.2467&0.2589&0.2518&0.3825&0.4405&0.3103&0.4486&0.5801&0.4114&0.3599\\
			&(0.0224)&(0.0235)&(0.0175)&(0.0175)&(0.0221)&(0.0173)&(0.0159)&(0.0161)&(0.0168)&(0.0149)&(0.0156)&(0.0159)\\
			\hline
			
			15&0.2651&0.2966&0.2509&0.2490&0.2843&0.3570&0.4603&0.2881&0.4144&0.3178&0.4097&0.3412\\
			&(0.0206)&(0.0215)&(0.0171)&(0.0173)&(0.0208)&(0.0173)&(0.0156)&(0.0163)&(0.0170)&(0.0183)&(0.0156)&(0.0159)\\
			\hline
			
			16&0.2296&0.2585&0.2318&0.3458&0.2680&0.3847&0.4194&0.3041&0.4145&0.3822&0.3706&0.3445\\
			&(0.0233)&(0.0240)&(0.0185)&(0.0160)&(0.0226)&(0.0179)&(0.0163)&(0.0167)&(0.0175)&(0.0179)&(0.0164)&(0.0162)\\
			\hline
			
			17&0.1980&0.2115&0.1944&0.1485&0.2611&0.2446&0.4595&0.2215&0.4229&0.2218&0.2670&0.2427\\
			&(0.0235)&(0.0252)&(0.0193)&(0.0224)&(0.0213)&(0.0209)&(0.0160)&(0.0182)&(0.0166)&(0.0232)&(0.0190)&(0.0188)\\
			\hline
			
			18&0.2007&0.2239&0.1786&0.1479&0.2468&0.2625&0.4848&0.1993&0.4072&NaN&NaN&NaN\\
			&(0.0234)&(0.0247)&(0.0199)&(0.0227)&(0.0217)&(0.0202)&(0.0158)&(0.0189)&(0.0169)&(NaN)&(NaN)&(NaN)\\
			\hline
			19&0.1928&0.3124&0.2001&0.1729&0.3059&0.2433&NaN&0.1665&0.4084&0.2678&0.2551&0.2298\\
			&(0.0258)&(0.0222)&(0.0202)&(0.0225)&(0.0214)&(0.0227)&(NaN)&(0.0220)&(0.0175)&(0.0217)&(0.0201)&(0.0200)\\
			\hline
			
			20&0.1920&0.2199&0.1876&0.1444&0.2420&0.2613&0.4187&0.2008&0.4009&0.2584&0.2592&0.2255\\
			&(0.0258)&(0.0265)&(0.0208)&(0.0241)&(0.0233)&(0.0217)&(0.0167)&(0.0200)&(0.0177)&(0.0221)&(0.0199)&(0.0199)\\
			\hline
			
			21&0.3257&0.2851&0.2579&0.3294&0.4767&0.6090&0.4404&0.3411&0.4144&0.4995&0.4600&0.4579\\
			&(0.0198)&(0.0231)&(0.0179)&(0.0163)&(0.0173)&(0.0154)&(0.0154)&(0.0159)&(0.0172)&(0.0163)&(0.0150)&(0.0149)\\
			\hline
			
			22&0.2818&0.2803&0.2657&0.2404&0.4073&0.4579&0.4407&0.2999&0.5427&0.4895&0.5175&0.4687\\
			&(0.0194)&(0.0214)&(0.0160)&(0.0172)&(0.0171)&(0.0157)&(0.0146)&(0.0153)&(0.0148)&(0.0157)&(0.0139)&(0.0143)\\
			\hline
			
			23&0.2994&0.3057&0.2703&0.3134&0.3426&0.4912&0.4511&0.3964&0.4488&0.7917&0.5295&0.5167\\
			&(0.0192)&(0.0209)&(0.0164)&(0.0155)&(0.0184)&(0.0157)&(0.0147)&(0.0141)&(0.0158)&(0.0140)&(0.0141)&(0.0142)\\
			\hline
			
			24&0.2915&0.3406&0.2984&0.2708&0.4359&0.4786&0.4219&0.3037&0.4241&0.7829&0.5281&0.5098\\
			&(0.0216)&(0.0217)&(0.0170)&(0.0181)&(0.0183)&(0.0169)&(0.0161)&(0.0171)&(0.0174)&(0.0150)&(0.0148)&(0.0147)\\
			\hline	
			
			25&0.3414&0.2773&0.2604&0.4262&0.3266&0.3939&0.4078&0.2519&0.5057&0.8064&0.4645&0.4960\\
			&(0.0193)&(0.0230)&(0.0174)&(0.0144)&(0.0198)&(0.0175)&(0.0156)&(0.0175)&(0.0155)&(0.0143)&(0.0149)&(0.0145)\\
			\hline	
				
			26&0.2952&0.3587&0.2600&0.2629&0.3224&0.5451&0.4192&0.2700&0.5214&0.4358&0.5040&0.4993\\
			&(0.0201)&(0.0199)&(0.0170)&(0.0172)&(0.0197)&(0.0154)&(0.0154)&(0.0168)&(0.0153)&(0.0168)&(0.0144)&(0.0144)\\
			\hline
			
			27&0.2843&0.3802&0.2992&0.2408&0.3750&0.3772&0.4377&0.2115&0.4591&0.3829&0.4387&0.3575\\
			&(0.0186)&(0.0178)&(0.0152)&(0.0169)&(0.0174)&(0.0164)&(0.0150)&(0.0177)&(0.0157)&(0.0163)&(0.0146)&(0.0152)\\
			\hline
			
			28&0.2568&0.3829&0.2780&0.2999&0.4873&0.3446&0.4541&0.2067&0.4115&0.3698&0.4152&0.3935\\
			&(0.0206)&(0.0186)&(0.0166)&(0.0160)&(0.0164)&(0.0177)&(0.0151)&(0.0187)&(0.0166)&(0.0174)&(0.0151)&(0.0150)\\
			\hline
			
			29&0.3090&0.4838&NaN&0.2490&0.4704&0.3834&0.4976&0.1925&0.5238&0.4670&0.4539&0.3602\\
			&(0.0180)&(0.0166)&(NaN)&(0.0164)&(0.0161)&(0.0162)&(0.0144)&(0.0182)&(0.0150)&(0.0151)&(0.0144)&(0.0152)\\
			\hline
			
			30&0.2674&0.3750&0.2569&0.2420&0.3766&0.3756&0.4564&0.2018&0.4188&0.4652&0.3871&0.3366\\
			&(0.0206)&(0.0191)&(0.0172)&(0.0179)&(0.0184)&(0.0174)&(0.0152)&(0.0194)&(0.0167)&(0.0162)&(0.0156)&(0.0159)\\
			\hline
			
			31&0.2747&0.2810&0.2852&0.2743&0.4059&0.4015&0.4331&0.2039&0.4369&0.3531&0.4292&0.3543\\
			&(0.0181)&(0.0192)&(0.0149)&(0.0152)&(0.0164)&(0.0153)&(0.0153)&(0.0173)&(0.0155)&(0.0162)&(0.0144)&(0.0150)\\
			\hline
			
			32&0.2814&0.3621&0.2885&0.2777&0.4317&0.3955&0.4195&0.2104&0.5036&0.3291&0.4665&0.3672\\
			&(0.0192)&(0.0186)&(0.0154)&(0.0161)&(0.0168)&(0.0163)&(0.0153)&(0.0181)&(0.0153)&(0.0178)&(0.0145)&(0.0152)\\
			\hline
			
			33&0.6782&0.3635&0.3431&0.2569&0.3591&0.4925&0.4426&0.3092&0.3879&0.3103&0.4241&0.4325\\
			&(0.0146)&(0.0196)&(0.0150)&(0.0174)&(0.0187)&(0.0158)&(0.0152)&(0.0162)&(0.0173)&(0.0193)&(0.0153)&(0.0148)\\
			\hline
			
			34&0.7057&0.7703&0.4077&0.3058&0.3802&0.4870&0.5149&0.3133&0.5033&0.4312&0.4717&0.4025\\
			&(0.0144)&(0.0144)&(0.0142)&(0.0162)&(0.0185)&(0.0159)&(0.0144)&(0.0162)&(0.0158)&(0.0168)&(0.0149)&(0.0150)\\
			\hline
			
			35&0.4773&0.5307&0.3696&0.2997&0.3601&0.3963&0.4544&0.2985&0.4507&1.0139&0.4542&0.4014\\
			&(0.0159)&(0.0160)&(0.0144)&(0.0162)&(0.0186)&(0.0169)&(0.0151)&(0.0161)&(0.0163)&(0.0135)&(0.0149)&(0.0149)\\
			\hline
			
			36&0.5652&0.5458&0.3376&0.3300&0.3686&0.4409&0.4465&0.2123&0.3741&0.3234&0.4014&0.4008\\
			&(0.0152)&(0.0160)&(0.0152)&(0.0156)&(0.0184)&(0.0165)&(0.0151)&(0.0188)&(0.0174)&(0.0189)&(0.0154)&(0.0150)\\
			\bottomrule
			\multicolumn{13}{l}{ Note: Table \ref{SiteinHebei} shows the specific city and site names with respect to site number.} 
	\end{tabular}}
\end{table}

\begin{figure}
	\centering
	\subfigure[Rank of average $\alpha$ in spring.]{
		\begin{minipage}[b]{0.45\textwidth}
			\includegraphics[width=1\textwidth]{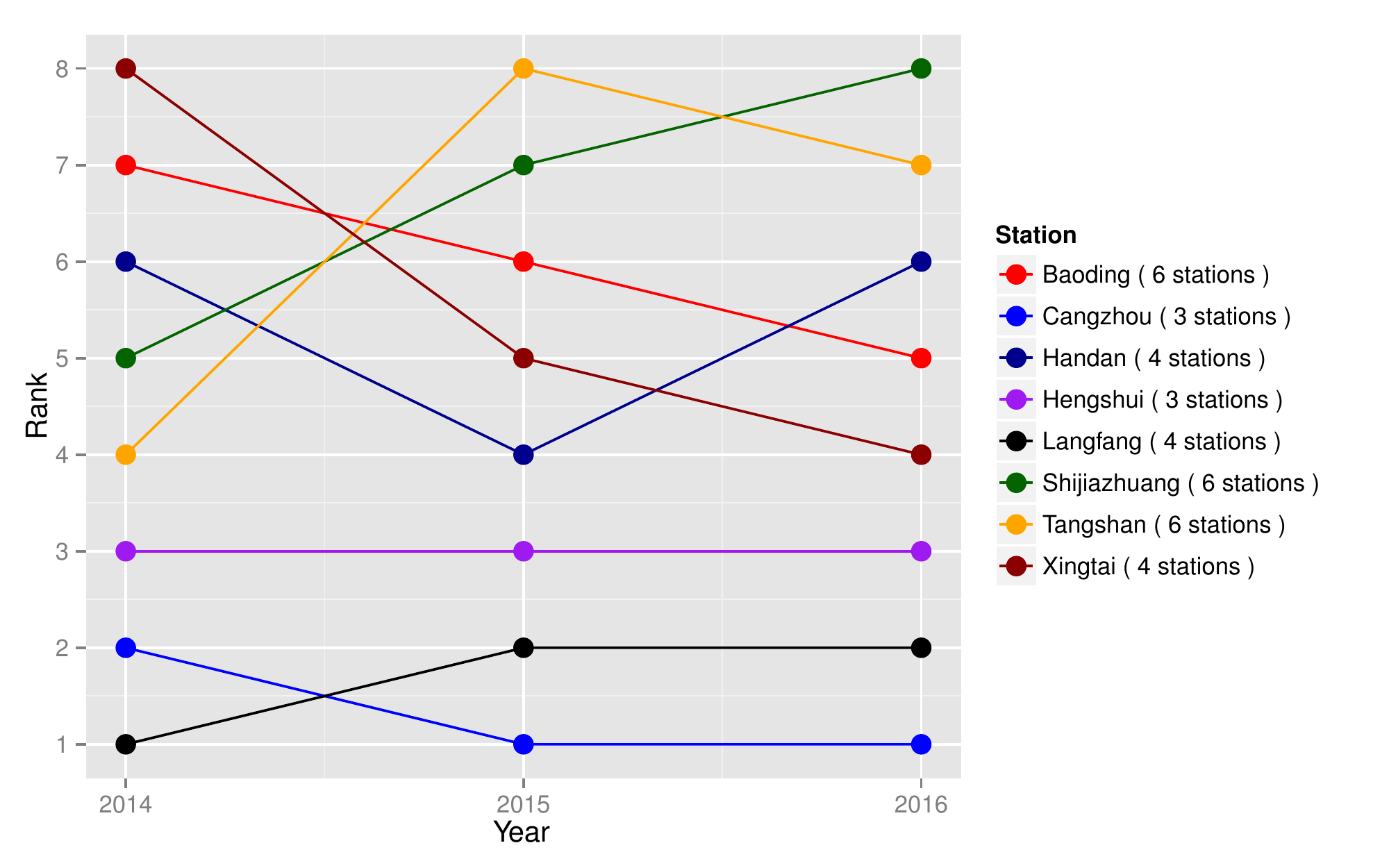}
		\end{minipage}
	}
	\subfigure[Rank of average $\alpha$ in summer.]{
		\begin{minipage}[b]{0.45\textwidth}
			\includegraphics[width=1\textwidth]{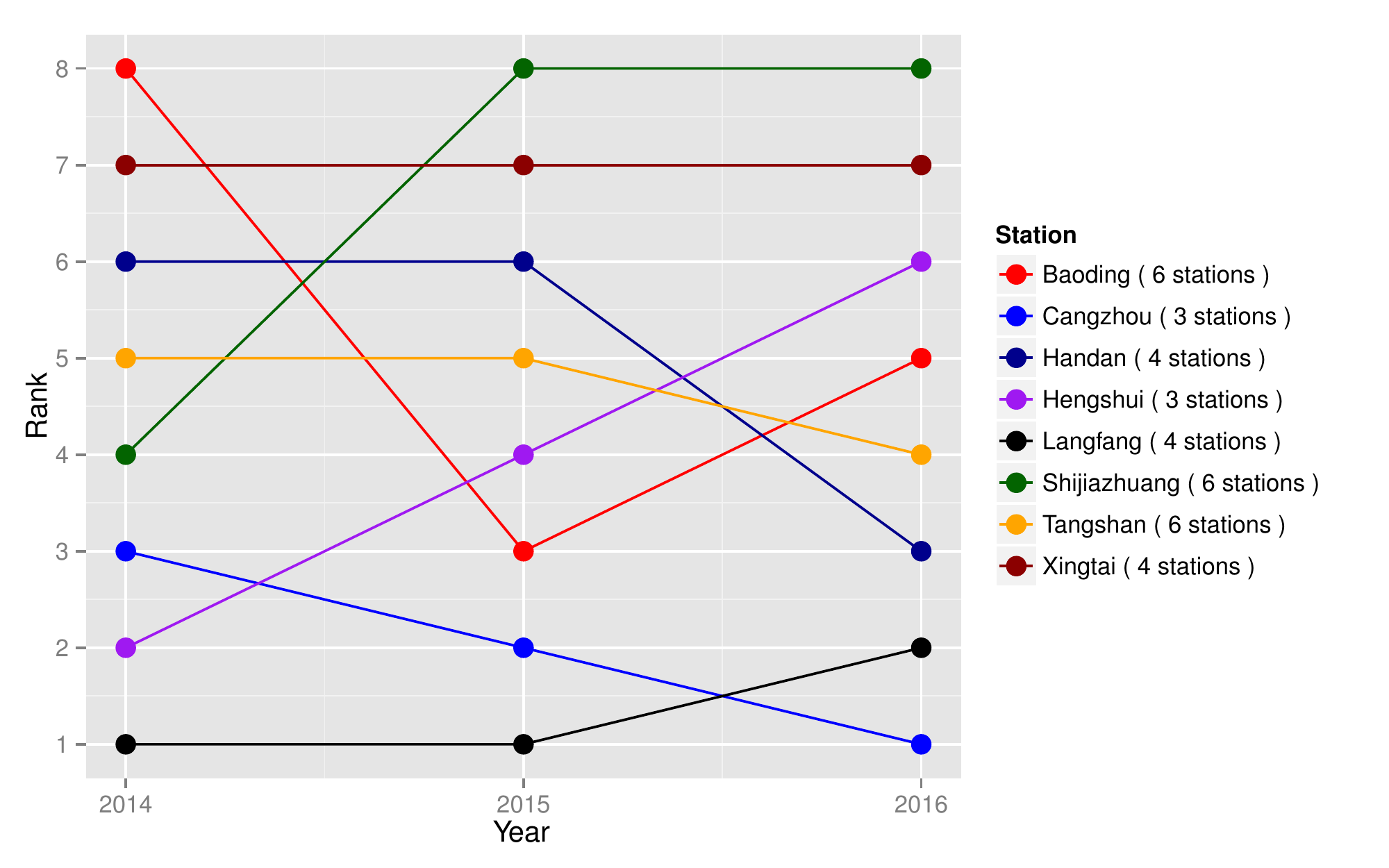}
		\end{minipage}
	}
	\subfigure[Rank of average $\alpha$ in autumn.]{
		\begin{minipage}[b]{0.45\textwidth}
			\includegraphics[width=1\textwidth]{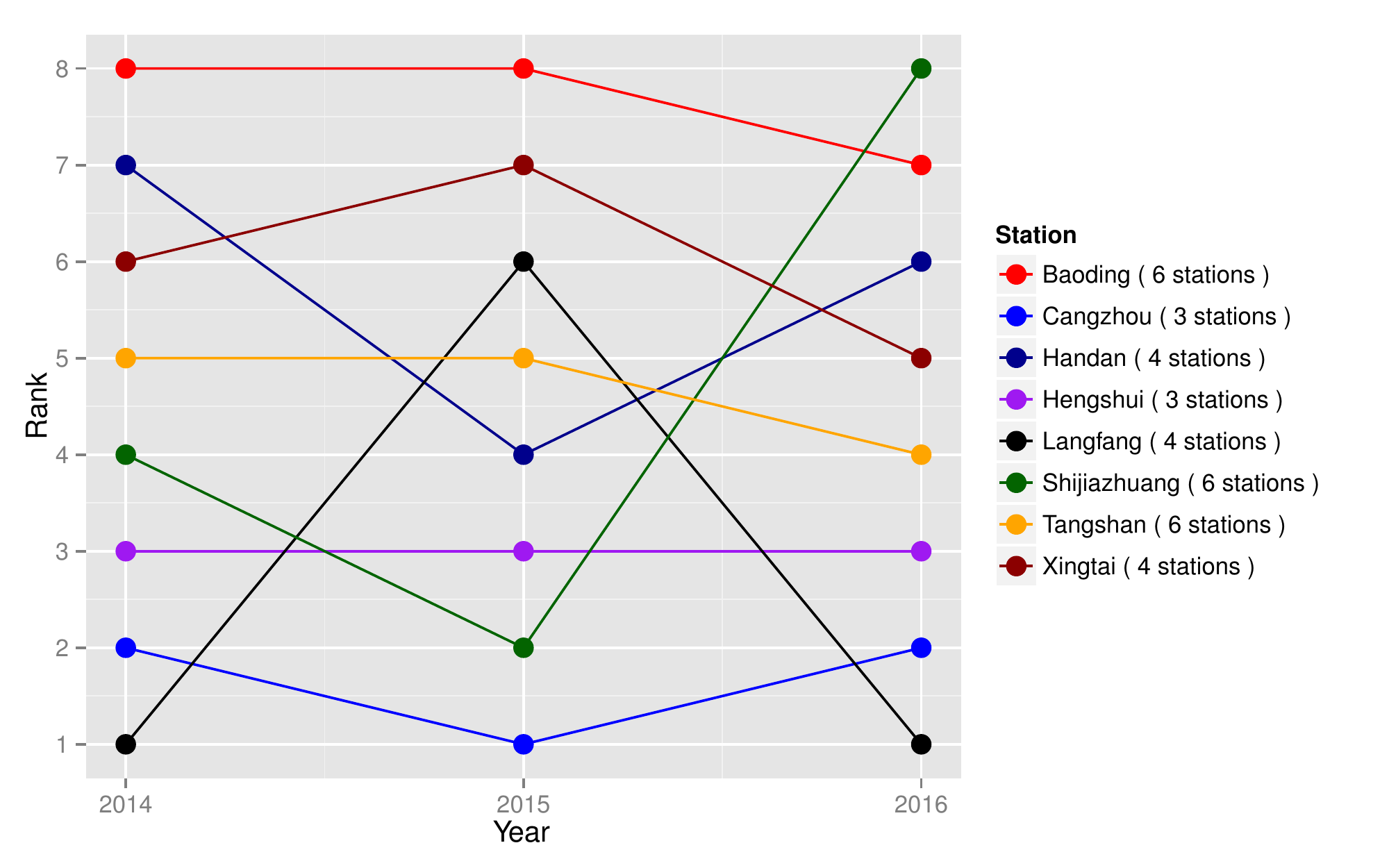}
		\end{minipage}
	}
	\subfigure[Rank of average $\alpha$ in winter.]{
		\begin{minipage}[b]{0.45\textwidth}
			\includegraphics[width=1\textwidth]{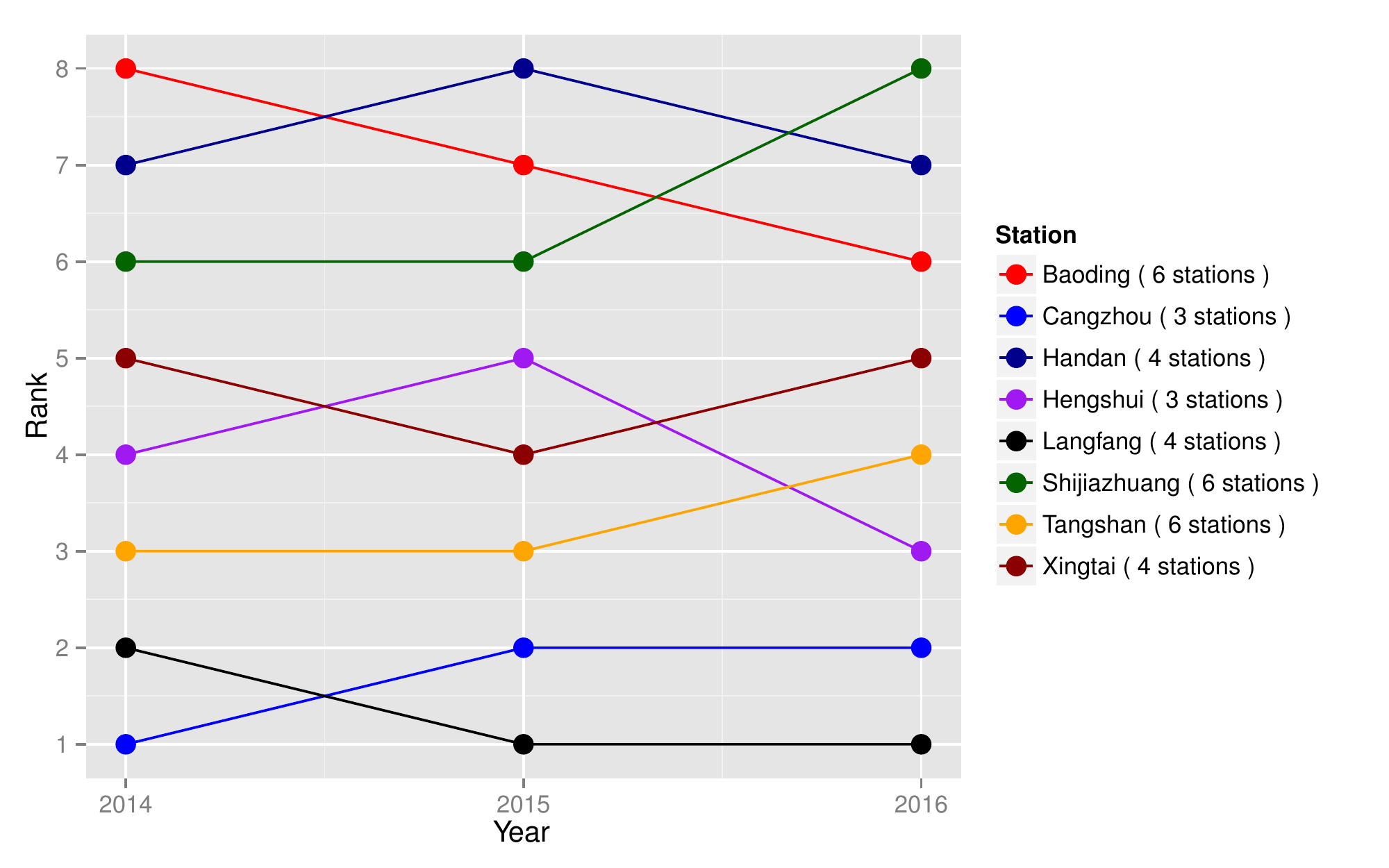}
		\end{minipage}
	}
	\caption{ \label{Seasonalphanew}The average $\alpha$ values for 8 cities, ranking from lowest to highest for each season.}	
\end{figure}

Since within the same city, the estimated $\alpha$ values of the 36 monitoring stations for 12 seasons bear close resemblance, we take the average of $\alpha$ values of the monitoring stations in the same city to facilitate comparison. Thus, we compute the average $\alpha$ values for the 8 cities in 12 seasons of 3 years. We rank them from 1 to 8, with 1 indicating the lowest average $\alpha$ value among the cities. Those with low $\alpha$ values are suspect of having biased readings. To assist visual inspection, we plot them in Figure \ref{Seasonalphanew} for the four seasons respectively. Figure \ref{Seasonalphanew} shows that the two cities--Cangzhou and Langfang, have consistently lower $\alpha$ values.

\begin{figure}
	\centering
	\includegraphics[width=12cm]{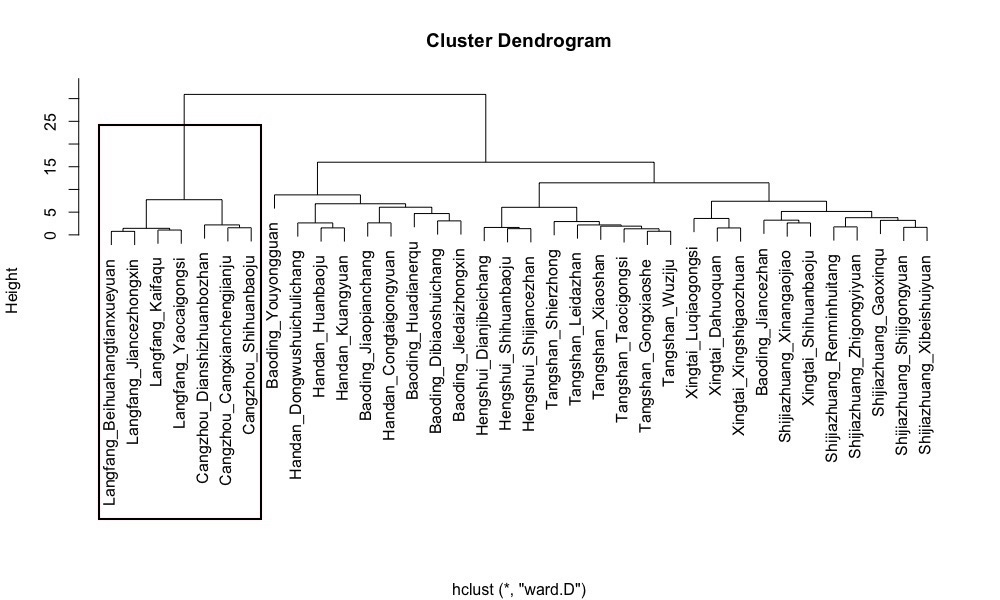}
	\caption{\label{clust} Cluster Result by Hierarchical Clustering. Monitoring stations in the box belong to Langfang and Cangzhou cities.}
\end{figure}

We use hierarchical clustering to classify 36 stations into several groups based on the distance matrix (Euclidean metric) of $\alpha$ values, and Figure \ref{clust} shows the classification tree using the hierarchical clustering algorithm with Ward's method. The figure clearly presents two categories. The mean and standard deviation of $\alpha$ values for the two groups are summarized in Table \ref{averagedalpha}. Again, we find that the seven monitoring stations in the low value category all belong to the two cities--Cangzhou and Langfang.

\begin{table}
	\caption{\label{averagedalpha} Mean and standard deviation of estimated values for the two groups.} 
	\centering
	\begin{tabular}{lcc}
		\toprule
		Group & Mean  &  Sd \\
		\hline
		1. stations in Langfang and Cangzhou cities &0.2548&0.0769\\
		\hline
		2. stations in other cities &0.4156&0.1578\\
		\bottomrule
	\end{tabular}	
\end{table}

\begin{figure}
	\centering
	\includegraphics[width=12cm]{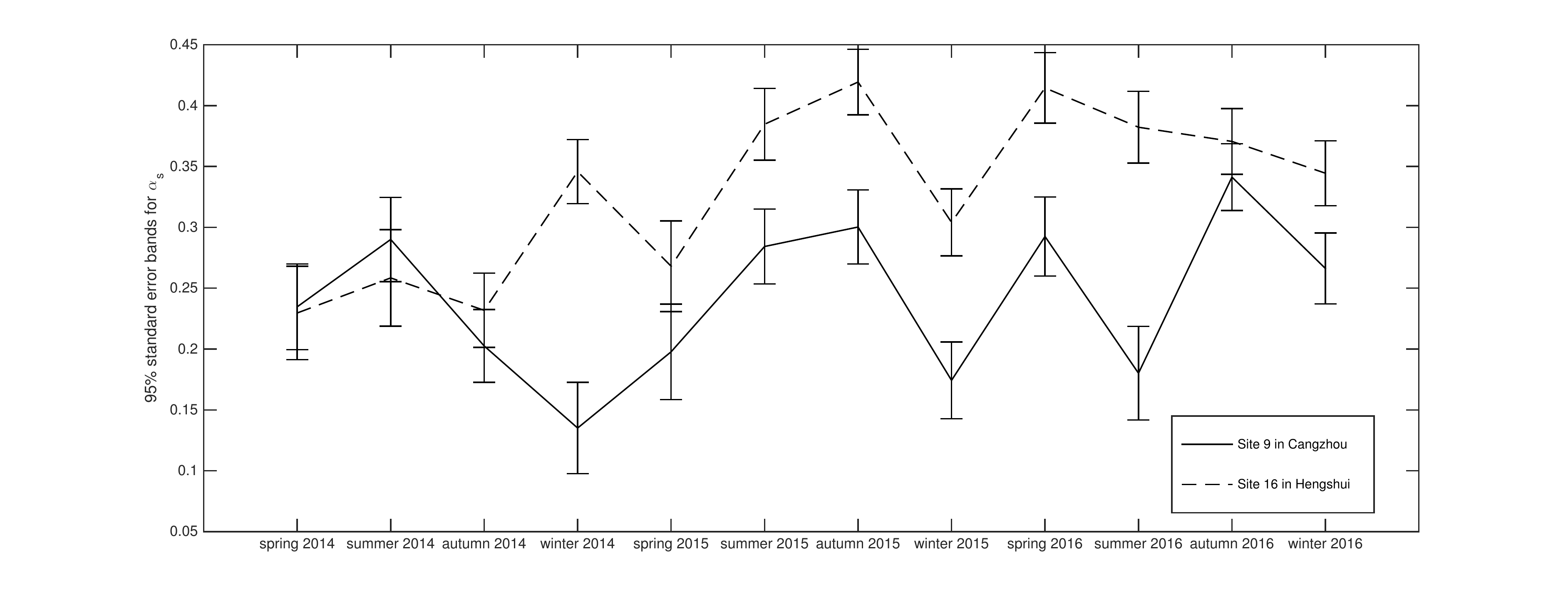}
	\caption{\label{ci_CH}Estimated $\alpha$ with its 95\% standard error bands for Site 9--Shihuanbaoju site in Cangzhou city, compared with Site 16--Shijiancezhan site in Hengshui city.}
\end{figure}

\begin{figure}
	\centering
	\includegraphics[width=12cm]{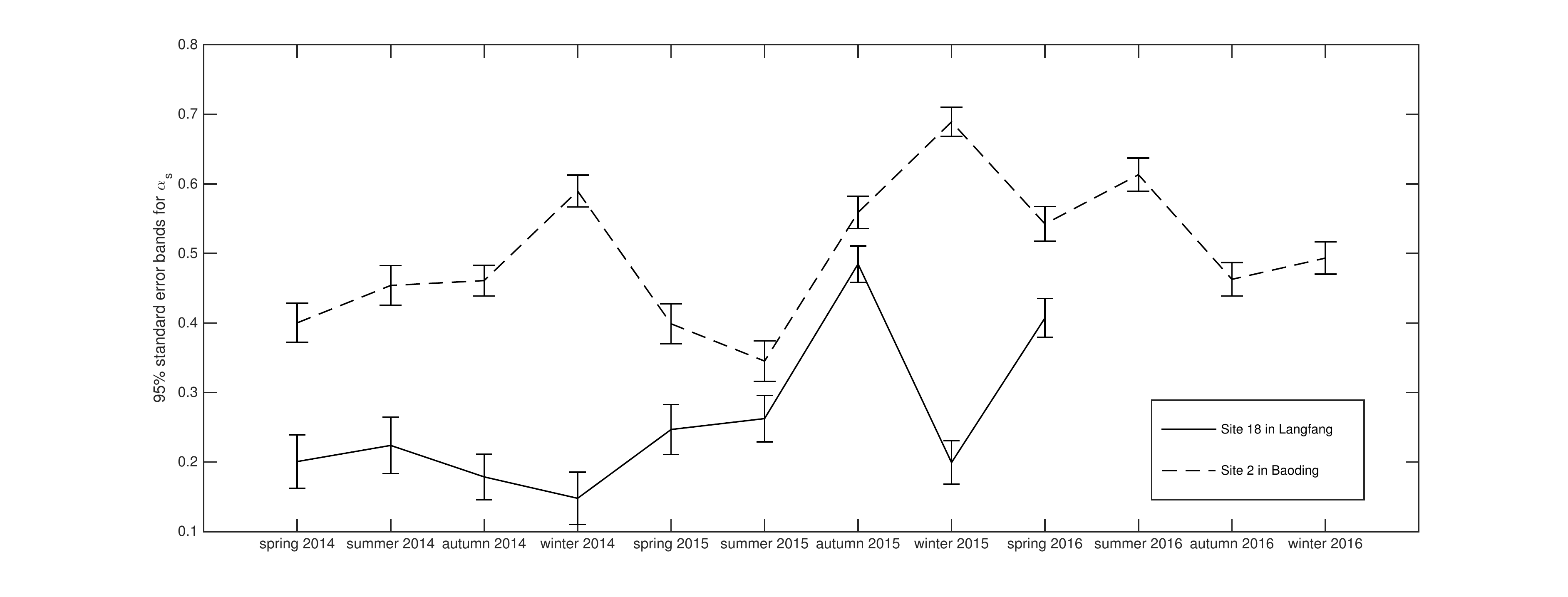}
	\caption{\label{ci_FB} Estimated $\alpha$ with its 95\% standard error bands for Site 18-Jiancezhongxin site in Langfang city, compared with Site 2--Huadianerqu site in Baoding city. $\alpha_{18}$ for Jiancezhongxin site is not estimated in summer, autumn, or winter 2016 due to the high missing rate (more than 20\%) of the observed $PM_{2.5}$.} 
\end{figure}

Moreover, we compare the $\alpha$ estimates by randomly choosing one site in the two identified cities with its nearest site in other city. For Cangzhou city, we choose the Shihuanbaoju site and its nearest site--the Shijiancezhan site in Hengshui city. The results for $\alpha$ estimate and its 95\% standard error bands of the two stations (see Figure \ref{ci_CH}) show a big difference emerges since late 2014, thus indicating that the readings from the site in Cangzhou may become biased since that time. In Figure \ref{ci_FB}, we compare the Jiancezhongxin site in Langfang city and its nearest site Huadianerqu in Baoding city, and find that the biggest difference occurs in winter heating period with extremely high $PM_{2.5}$, which indicates the $PM_{2.5}$ readings in Langfang site may be biased when heavy air pollution comes up. 

To further investigate these two stations, we examine their pollution data through comparing both $PM_{2.5}$ and $PM_{10}$ levels. We consider the concentration ratio of $PM_{2.5}$ to $PM_{10}$. If the $PM_{2.5}$ concentration is much lower than that of $PM_{10}$, it may indicate biased readings. After checking the data, we do find some suspicious cases. For example, in Table \ref{Ratio1}, we find that the average $PM_{2.5}$ from Shihuanbaoju, Cangzhou, was much lower than that from Shijiancezhan, Hengshui, whereas the average $PM_{10}$ was much closer in the same period from February 8 to 10, 2016. In Table \ref{Ratio2}, from June 19 to 20, 2014, the average $PM_{10}$ from Jiancezhongxin site in Langfang city and its nearest site–the Huadianerqu site in Baoding city– was quite high, while the average $PM_{2.5}$ from Jiancezhongxin, Langfang was much less than that of Baoding. The big difference in the average ratio of $PM_{2.5}$ to $PM_{10}$ again provides evidence that the readings of $PM_{2.5}$ in Langfang city may be biased. 

Biased reading of $PM_{2.5}$ in Langfang is also mentioned in the Air Quality Assessment Report (4) --Regional Pollution Situation Assessment of Beijing, Tianjin and Hebei in 2013-2016, published by Peking University. In the report, the four stations in Langfang are compared to the two geographically nearest monitoring stations in the south of Beijing. In the four seasons of 2016, the quarterly average $PM_{2.5}$ concentration of the two stations in Beijing was much higher than the quarterly average of the four stations in Langfang city, which also hints that there may exist biased reading in Langfang. 

\begin{table}
	\caption{\label{Ratio1} Mean of $PM_{10}$, $PM_{2.5}$ and their ratio, with standard deviation in parentheses for the two stations from February 8 to 10, 2016.}
	\centering
	\begin{tabular}{@{} *{4}{c} @{}} 
		\toprule
		& $PM_{2.5}$& $PM_{10}$ & Ratio ($PM_{2.5}/PM_{10}$) \\
		\hline
		Shihuanbaoju in Cangzhou city  &  60  &  152 &  38\%
		\\
		(Site 9)  & (35) &  (80) &  (9\%) \\
		\hline
		Shijiancezhan in Hengshui city  & 134  & 187  & 70\% \\
		(Nearest site, Site 16)  & (75) & (97) & (10\%)  \\
		\bottomrule
	\end{tabular}
\end{table}

\begin{table}
	\caption{\label{Ratio2} Mean of $PM_{10}$, $PM_{2.5}$ and their ratio, with standard deviation in parentheses for the two stations from June 19 to 20, 2014.}
	\centering
	\begin{tabular}{@{} *{4}{c} @{}}
		\toprule
		& $PM_{2.5}$& $PM_{10}$ & Ratio ($PM_{2.5}/PM_{10}$) \\
		\hline
		Jiancezhongxin in Langfang city & 104 & 226 & 46\% \\
		(site 18)& (56) & (124) & (4\%) \\
		\hline
		Huadianerqu in Baoding city & 160 & 242 & 69\% \\
		(Nearest site, site 2) & (76) & (113) & (13\%) \\
		\bottomrule
	\end{tabular}
\end{table}

\section{Discussions}
In this paper, we propose a hidden dynamic geostatistical calibration model to detect biased readings from monitoring stations. The method has several advantages. First, the station-specific calibration component $\alpha_s$ of a spatial-temporal model help us automatically find out stations with biased readings. We further investigate those detected stations through the comparison of calibration component $\alpha_s$ between monitoring stations, and from another perspective the concentration ratio of $PM_{2.5}$ to $PM_{10}$. Second, as a hierarchical spatial temporal statistical model, the model is flexible enough to handle data with missing values and latent variable while capturing spatio-temporal dynamics. 

Moreover, the method is efficient in discovering relational patterns from chemical reaction, meteorological, and geographical factors on the formation of $PM_{2.5}$. Among the four gaseous precursors, $NO_2$ contributes to the largest influence on $PM_{2.5}$ on average. Given that motor vehicle exhaust fumes are the main source of $NO_2$, this result indicates a more specific traffic-related type of $PM_{2.5}$ pollution (e.g. \cite{richter2005increase} and \cite{zhang2007trend}). Motor vehicles data from National Bureau of Statistics of the People’s Republic of China shows a phenomenal surge in the number and use of motor vehicles in Hebei province. For example, the number of possessions of private cars in Hebei was 719 million in 2014, and it increased to 1143 million (nearly 60\%) in 2016. So we suggest urban traffic should be one focus of policy action if air quality is to be improved. In the heating period (autumn and winter), the impact from $CO$ exceeds that of $NO_2$. $CO$ can come from fossil fuel combustions as well as motor vehicle emission. This result hints that fossil fuel combustions can largely exacerbate air pollution in addition to the increasing motor vehicle. Nearly all the meteorological variables have significant impact on $PM_{2.5}$, and dew point is the most influential meteorological factor. Beside, the geographic variable, distance to mountain, negatively relates to $PM_{2.5}$ concentration, as it is easier to accumulate the polluted air near the mountain. 


Finally, confronted with the great air-pollution challenges in China, we appeal that all Chinese cities should commit to ensuring the accuracy and reliability of air pollution data and further reducing $PM_{2.5}$ levels.

\section*{Acknowledgements}
This research is funded by a China's National Key Research Special Program Grants 2016YFC0207701 and 2016YFC0207702. We would also like to thank the support from Center for Statistical Science in Peking University, and the Key Laboratory of Mathematical Economics and Quantitative Finance (Peking University), Ministry of Education.

\bibliographystyle{chicago}

\end{document}